%% file: main.tex
\documentclass[lettersize,journal]{IEEEtran}
\usepackage{amsmath,amsfonts}
\usepackage{algorithmic}
\usepackage{algorithm}
\usepackage{array}
\usepackage[caption=false,font=normalsize,labelfont=sf,textfont=sf]{subfig}
\usepackage{textcomp}
\usepackage{stfloats}
\usepackage{url}
\usepackage{verbatim}
\usepackage{graphicx}
\usepackage{cite}
\usepackage{xspace}
\usepackage{listings}
\usepackage[frozencache]{minted}
\usepackage{tabularx}
\usepackage{graphicx}
\usepackage{booktabs}
\usepackage{todonotes}
\usepackage{caption}
\usepackage{multirow}
\usepackage{subfig}
\usepackage[dvipsnames]{xcolor}
\usepackage{tikz}
\usepackage{hyperref}
\usetikzlibrary{shapes,arrows,positioning}
\newcommand{\ourtool}{\textsc{OMPilot}\xspace}
\newcommand{\ourmetric}{\textsc{OMPBLEU}\xspace}

\newcommand{\greenup}{\textcolor{green}{$\uparrow$}}
\newcommand{\reddown}{\textcolor{red}{$\downarrow$}}
\usepackage[most]{tcolorbox}
\newtcolorbox{chatbox}[2][]{
  colback=white,
  colframe=black!50,
  boxrule=0.5pt,
  arc=1pt,
  auto outer arc,
  fonttitle=\bfseries,
  title=#2,
  #1
}
\begin{document}

\title{\ourtool: Harnessing Transformer Models for Auto Parallelization to Shared Memory Computing Paradigms}

\author{%
Arijit Bhattacharjee\IEEEauthorrefmark{1},
Ali TehraniJamsaz\IEEEauthorrefmark{2},
Le Chen\IEEEauthorrefmark{3},
Niranjan Hasabnis\IEEEauthorrefmark{4},
Mihai Capota\IEEEauthorrefmark{5},
Nesreen K.~Ahmed\IEEEauthorrefmark{6},
Ali Jannesari\IEEEauthorrefmark{1}%
\thanks{\IEEEauthorrefmark{1} Department of Computer Science, Iowa State University, Ames, IA, USA.}%
\thanks{\IEEEauthorrefmark{2} Mako AI, USA.}%
\thanks{\IEEEauthorrefmark{3} Argonne National Laboratory, USA.}%
\thanks{\IEEEauthorrefmark{4} Code Metal, USA.}%
\thanks{\IEEEauthorrefmark{5} Intel Labs, USA.}%
\thanks{\IEEEauthorrefmark{6} Cisco Outshift, USA.}%
}



\maketitle

\begin{abstract}
Recent advances in large language models (LLMs) have significantly accelerated progress in code translation, enabling more accurate and efficient transformation across programming languages. While originally developed for natural language processing, LLMs have shown strong capabilities in modeling programming language syntax and semantics, outperforming traditional rule-based systems in both accuracy and flexibility. These models have streamlined cross-language conversion, reduced development overhead, and accelerated legacy code migration. 
In this paper, we introduce \ourtool, a novel domain-specific encoder-decoder transformer tailored for translating C++ code into OpenMP, enabling effective shared-memory parallelization. 
\ourtool leverages custom pre-training objectives that incorporate the semantics of parallel constructs and combines both unsupervised and supervised learning strategies to improve code translation robustness.
% We tailor specific pre-training objectives that are crucial to capture the code semantics and structures of parallel constructs. We implement a combination of unsupervised and supervised training objectives to enhance our model's capability. 
Unlike previous work that focused primarily on loop-level transformations, \ourtool operates at the function level to capture a wider semantic context. To evaluate our approach, we propose \ourmetric, a novel composite metric specifically crafted to assess the correctness and quality of OpenMP parallel constructs, addressing limitations in conventional translation metrics.  
\end{abstract}

\input{sections/1-introduction}

\input{sections/2-background}
\input{sections/3-OMP}
\input{sections/4-OMPBLEU}
\input{sections/5-experimental_setup}
\input{sections/6-results}
\input{sections/7-conclusion_future}

\bibliographystyle{IEEEtran}
\bibliography{main.bbl}
\end{document}

%% file: sections/1-introduction.tex
\section{Introduction}
% The end of Dennard scaling led to multi-core processors and the rise of shared-memory parallelism, where multiple threads or processes access the same memory. % This shift spurred the development of parallel programming models like OpenMP~\cite{dagum1998openmp}, Cilk~\cite{frigo1998cilk}, and TBB~\cite{pheatt2008intel}. % OpenMP, favored for its open nature, uses compiler directives (\emph{pragmas}) to convert serial code into parallel code that can exploit multi-core architectures. Writing parallel programs manually was challenging because programmers not only had to master OpenMP pragmas but also contend with factors like loop-carried dependencies and estimating potential performance gains through parallelization. This complexity spurred the development of \emph{automatic parallelization}—tools designed to automatically transform serial code into its parallel counterpart.
The end of Dennard scaling marked a shift toward multi-core processors\cite{bohr200930} and highlighted the growing importance of shared-memory parallelism, where multiple threads or processes access a common memory space. This architectural shift led to the development of parallel programming models such as OpenMP~\cite{dagum1998openmp}, Cilk~\cite{frigo1998cilk}, and Intel TBB~\cite{pheatt2008intel}. Among these, OpenMP emerged as a widely adopted standard due to its open nature and ease of use via compiler directives (\emph{pragmas}) that facilitate the transformation of serial code into parallel code capable of exploiting multi-core hardware. However, writing efficient parallel programs remains a challenging task. Developers must not only master the intricacies of OpenMP pragmas but also reason about loop-carried dependencies, synchronization overhead, and the potential performance benefits of parallelization. These complexities have motivated the development of \emph{automatic parallelization} tools, which aim to convert sequential programs into parallel implementations. 

Automatic parallelization tools generally fall into two broad categories: formal tools and AI-based tools. Formal tools are static, rule-driven systems, including auto-parallelizing compilers (e.g., GCC, LLVM, ICC) that rely on loop-carried dependence analysis to parallelize \texttt{for} loops. In addition, they also include source-to-source transformation tools such as Cetus~\cite{lee2003cetus} and Par4All~\cite{amini2012par4all}, which transform sequential code into its parallel equivalent. In contrast, AI-based tools leverage large-scale code repositories (e.g., GitHub) to learn parallelization patterns directly from raw program text~\cite{chen2024OMPGPT, kadosh2023pragformer, mahmud2023autoparllm}. By bypassing explicit syntax parsing and handcrafted transformation rules, these approaches promise greater flexibility, improved generalization, and reduced manual effort in both development and maintenance. 
Although both formal and AI-based auto-parallelization tools have shown promise, several studies have highlighted their limitations~\cite{harel2020source, nichols2024large, chen2024landscape}. Formal tools, in particular, tend to be overly conservative, often missing viable parallelization opportunities. Their reliance on heuristic-based hardware cost models to assess loop parallelization potential can lead to suboptimal or even performance-degrading transformations. Additionally, these tools require continuous manual maintenance, which can delay support for new OpenMP pragmas and evolving language features. AI-based tools also face significant challenges, including the limited availability of high-quality OpenMP training data, the risk of aggressive parallelization without adequate correctness guarantees, and the substantial computational resources required for training and fine-tuning~\cite{nichols2024large}. 

%Our analysis of existing auto-parallelization tools revealed several key shortcomings. In particular, we identify two critical limitations of AI-based approaches\footnote{We discuss limitations of formal tools in detail in the experimental evaluation section.}. First, we observe that slight variations in natural language (NL) prompts can lead to substantially different parallelized codes. For example, Listing 1 illustrates two divergent outputs from the o3-mini model for the same serial input, differing only in the NL prompt. This raises an important question: \emph{Is natural language even necessary for AI-based auto-parallelization?} Removing the NL layer could potentially lead to smaller, more efficient models with better controllability.
Our analysis of existing auto-parallelization tools revealed key shortcomings in AI-based approaches\footnote{Limitations of formal tools are discussed in the experimental evaluation section.}. Slight variations in natural language (NL) prompts can lead to vastly different parallelized outputs, for instance, Listing 1 shows two divergent outputs from the o3-mini model for the same input, differing only in the NL prompt. This observation prompts us to ask: \emph{Is natural language necessary for AI-based auto-parallelization?} Removing the NL layer could lead to smaller, more efficient models with improved controllability.\\
%Second, we find that widely used evaluation metrics such as BLEU~\cite{papineni2002bleu} and CodeBLEU~\cite{ren2020codebleu} are insufficient for assessing the correctness of parallelized code. While BLEU and CodeBLEU are designed to measure the similarity in natural and programming languages, they are not well-suited to capturing the structural and semantic correctness of parallel programming constructs like OpenMP directives. For instance, \autoref{fig:ompbleu_motivation} compares a ground-truth OpenMP implementation with a model-generated version. Despite high BLEU and CodeBLEU scores, the generated output contains incorrect OpenMP pragmas that alter the program’s semantics—highlighting the inadequacy of general-purpose metrics in this context.
Second, our findings indicate that conventional evaluation metrics such as BLEU~\cite{papineni2002bleu} and CodeBLEU~\cite{ren2020codebleu} fall short in assessing the correctness of parallelized code. Although these metrics effectively measure textual similarity in natural and programming languages, they do not capture the structural and semantic integrity of parallel constructs like OpenMP directives. As illustrated in \autoref{fig:ompbleu_motivation}, even when high BLEU and CodeBLEU scores are achieved, the model-generated output can contain flawed OpenMP pragmas that alter the intended program semantics, underscoring the need for more specialized evaluation metrics in this context.
\begin{figure}[!htb]
  \centering
  \begin{minipage}[t]{0.48\linewidth}
    \centering
    \begin{minted}[breaklines,frame=lines,framesep=1mm,baselinestretch=1.0,bgcolor=white,fontsize=\footnotesize]{c}
/* GROUND TRUTH */

#define N 100
int a[N][N];
int main(void) {
  int i, j; long sum=0;
  #pragma omp parallel for collapse(2) private(i,j) reduction(+:sum) schedule(static)
  for (i = 0; i < N; ++i) {
    for (j = 0; j < N; ++j) {
      a[i][j] += 1; 
      sum += a[i][j];
    }
  }
  return 0;
}
    \end{minted}
    \subfloat{Ground truth parallel code}
  \end{minipage}
\hfill
\begin{minipage}[t]{0.48\linewidth}
\centering
\begin{minted}[breaklines,frame=lines,framesep=1mm,baselinestretch=1.0,bgcolor=white,fontsize=\footnotesize]{c}
/* INCORRECT CODE */

#define N 100
int a[N][N];
int main(void) {
  int i, j; long sum=0;
  #pragma omp parallel for collapse(2) private(i,j)
  for (i = 0; i < N; ++i) {
    for (j = 0; j < N; ++j) {
      a[i][j] += 1;
      sum += a[i][j];
    }
  }
  return 0;
}

\end{minted}
\subfloat{Incorrect parallel code}
\end{minipage}
\caption{Comparison between BLEU/CodeBLEU and \ourmetric{} on the ground‑truth and the generated OpenMP parallel code. Even though the generated code omits the \texttt{reduction(+:sum) schedule(static)} clauses, both BLEU/CodeBLEU scores are high, 90.82/98.78. \ourmetric{} score for the pair is 57.49.}
\label{fig:ompbleu_motivation}
\end{figure}
\begin{tcolorbox}[
  enhanced, 
  colback=black!5,
  colframe=black!70,
  halign title=center,
  title= Listing 1: OpenAI o3-mini Prompt Sensitivity,
  boxrule=0.7pt,
  left=1mm, right=1mm, top=1mm, bottom=1mm,scale=0.6
]
\centering
% ─────────────────────────────────────────────────────────────────────────────
% Top row: two prompts side by side
\begin{minipage}[t]{0.48\linewidth}
  \textbf{Prompt:}
  \medskip

  \begin{minted}[fontsize=\scriptsize,breaklines,frame=lines,framesep=1mm]{c}
int main()
{
    int i, j;
    for (i = 0; i < N; i++) {
        for (j = 0; j < N; j++) {
            a[i][j] = a[i][j] + 1;
        }
    }
    return 0;
}

parallelize it with openmp

  \end{minted}
\end{minipage}\hfill
\begin{minipage}[t]{0.48\linewidth}
  \textbf{Prompt:}
  \medskip

  \begin{minted}[fontsize=\scriptsize,breaklines,frame=lines,framesep=1mm]{c}
int main()
{
    int i, j;
    for (i = 0; i < N; i++) {
        for (j = 0; j < N; j++) {
            a[i][j] = a[i][j] + 1;
        }
    }
    return 0;
}
can you look at this C++ code and give me a parallelized version with openmp
  \end{minted}
\end{minipage}

\vspace{1em}

% Bottom row: two responses side by side
\begin{minipage}[t]{0.48\linewidth}
  \textbf{Response:}
  \medskip

  \begin{minted}[fontsize=\scriptsize,breaklines,frame=lines,framesep=1mm]{c}
#include <omp.h>
#define N 1000  /* or whatever size */
int a[N][N];

int main(void) {
    int i, j;
    #pragma omp parallel for private(j)
    for (i = 0; i < N; i++) {
        for (j = 0; j < N; j++) {
            a[i][j] += 1;
        }
    }

    return 0;
}

  \end{minted}
\end{minipage}\hfill
\begin{minipage}[t]{0.48\linewidth}
  \textbf{Response:}
  \medskip

  \begin{minted}[fontsize=\scriptsize,breaklines,frame=lines,framesep=1mm]{c}
#include <omp.h>
#include <iostream>
const int N = /* your size */;
int a[N][N];

int main() {
    for (int i = 0; i < N; ++i)
        for (int j = 0; j < N; ++j)
            a[i][j] = 0;

    #pragma omp parallel for collapse(2) schedule(static)
    for (int i = 0; i < N; ++i) {
        for (int j = 0; j < N; ++j) {
            a[i][j] += 1;
        }
    }
  \end{minted}
\end{minipage}

\end{tcolorbox}
Motivated by current limitations, we introduce \ourtool{}, a lightweight AI model for OpenMP-based parallelization, and \ourmetric{}, a novel metric for evaluating OpenMP code. By removing natural language prompts, \ourtool{} simplifies the process, enhances inference efficiency, and employs a unique loss function to better generate accurate parallel pragmas. Meanwhile, \ourmetric{} captures the syntactic and semantic nuances of OpenMP directives, offering a more reliable assessment than BLEU or CodeBLEU. Experimental results show that \ourtool{} produces more accurate, efficient parallel code, with \ourmetric{} strongly correlating with semantic correctness.
%most of the tools, if not all, automatically parallelize \texttt{for} loops only. This limitation stems from \textbf{What}. When we subjected these tools to automatic parallelization of C++ code from StackV2, we found that these tools missed X\% of the parallelizable code elements such as block statements in C++. As an example, Figure\textbf{X} shows the input C++ code we fed to tool \textbf{T}; Figure\textbf{Z} shows the parallelization output of the tool; Figure\textbf{Z} shows the places where the tool missed the parallelization opportunity. \arijit{See if you can find this example.}

\textbf{Contributions.} This paper makes the following contributions:

\begin{itemize}

\item \textbf{\ourtool{}} introduces a lightweight (0.8B) and efficient AI model for OpenMP-based auto-parallelization, offering several advantages over existing approaches: 
 \begin{itemize}

    \item \textbf{Weighted Token Cross-Entropy Loss function} for specifically targeting OpenMP reserved keyword tokens
    \item \textbf{Syntax Structure Annotation:}To emphasize on the correct placement of the OpenMP clauses w.r.t the surrounding context.
    \item \textbf{Broader Support for OpenMP Clauses:} While most existing tools focus on \texttt{for}-loop parallelization~\footnote{Loops often dominate runtime in compute-intensive applications.}, \ourtool{} supports both loop-level and block-level parallelism. This enables support for a wider range of OpenMP clauses than prior work.
    % \item \textbf{Broader support for OpenMP clauses}: Most of the existing auto-parallelization tools focus on parallelizing \texttt{for} loops in high-level languages~\footnote{Because programs spent majority of their time in loops}. \ourtool{}, on the other hand, supports block-level parallelism, in addition to \texttt{for} loop level parallelism. This conscious choice has led \ourtool{} to support a broad set of OpenMP clauses, considerably more than those supported by existing tools.
    \item \textbf{Efficiency:} By removing the dependency on natural language prompts, \ourtool{} reduces model size and improves training and inference efficiency compared to existing AI-based models.
    % \item \textbf{Efficient}: Eliminating the need for natural language input to \ourtool{} helps us ensure that \ourtool{} is considerably smaller and thus efficient to train/infer than existing AI-based models for auto-parallelization.
 \end{itemize}
    
    \item \textbf{\ourmetric{}} is a novel evaluation metric specifically designed for OpenMP. By incorporating both the syntax and semantics of OpenMP pragma directives, \ourmetric{} captures key elements of parallel programs, offering a more reliable assessment of correctness than general-purpose metrics like BLEU and CodeBLEU.
 % \item \textbf{\ourmetric{}}: Our experimental analysis reveals that existing metrics for OpenMP-based parallel code have limitations. We improve upon these limitations by developing a novel metric that exploits the key elements of parallel programs, in addition to OpenMP pragmas, to measure syntactic and semantic validity of generated parallel programs.

 % \item \textbf{Experimental results}: Our experimental results parallel programs generated by \ourtool{} are more accurate (by X) and efficient (by how much) over the best existing auto-parallelization tool. Moreover, the results also demonstrate that \ourmetric{} measures syntactic and semantic validity of generated parallel programs much better than existing metrics.

 \item \textbf{Experimental Evaluation:} We conduct comprehensive evaluations showing that \ourtool{} generates parallel programs with a higher OMPBLEU score (by 9.61\%) and greater efficiency (28$\times$ faster inference) compared to the next best among existing LLMs. We also show that \ourmetric{} correlates more strongly with semantic correctness, outperforming prior metrics in evaluating parallel code quality. We also show our evaluation on a real world benchmark: XSBench\cite{Tramm:wy}.
\end{itemize}

    %\item so far talking about llm based tools refer to existing OMP-LLM papers for their support for different clauses; some llms have closed-source training set; starcoderv2 is trained on StackV2 (same as us) but is very large because of number of languages that it supports. (2) certain clauses are not loop-level, existing tools dont support such clauses because they only work at loop-level

%% file: sections/2-background.tex
\section{Background and Related Works}
This section provides a comprehensive background and overview of past literature relevant to our work. It reviews key developments in auto parallelization and code translation, detailing earlier rule-based approaches alongside more recent machine learning-driven methods.
\subsection{Rule-based Auto Parallelization}

Rule-based auto parallelizers, such as AutoPar~\cite{liao2010semantic}, Intel's ICC Classic compiler, and Cetus~\cite{lee2003cetus}, are early tools designed to automatically transform sequential code into parallel code. They rely on predefined rules and static analysis to examine data dependencies, control flow, and loop structures, converting serial loops into parallel constructs. Additionally, these tools typically incorporate a cost model to assess performance trade-offs and choose the most efficient parallelization strategies. AutoPar uses dependency graphs for loop-level parallelization, the ICC compiler combines rule-based transformations with multi-core optimizations, and Cetus employs source-to-source transformations. However, their reliance on fixed rules limits their adaptability to modern, complex software, paving the way for more dynamic, machine learning-based methods. 
%subsection{Neural Machine Translators}
%Code translation builds upon advances in natural language machine translation (NMT). Early work relied on statistical machine translation (SMT) techniques~\cite{10.5555/972470.972474,10.3115/1073445.1073462} that used phrase-based models from parallel corpora, but these approaches often struggled with long-range dependencies and fluency. The advent of neural machine translation, particularly sequence-to-sequence models with attention mechanisms~\cite{bahdanau2016neuralmachinetranslationjointly}, enabled end-to-end mapping of input to output sequences, capturing complex relationships. The Transformer architecture~\cite{vaswani2023attentionneed}, with its self-attention mechanism, further improved performance through parallelization and better handling of long-range dependencies. Our work leverages the Transformer to capture the nuances and complexities inherent in programming languages.
\subsection{LLMs for Code}

Large language models are now central to code generation and translation. Strong general-purpose coders like DeepSeek-Coder~\cite{zhu2024deepseek} and Qwen2.5-Coder~\cite{hui2024qwen2} provide robust baselines, which can be further improved via fine-tuning e.g., Code Alpaca~\cite{codealpaca} on LLaMA or with parameter-efficient approaches such as Astraios~\cite{zhuo2024astraios}. Moreover, advances in prompt design~\cite{taherkhani2024epiccosteffectivesearchbasedprompt} have yielded notable gains in output quality. Building on these developments, \ourtool introduces new metrics and techniques specifically aimed at improving C++→OpenMP translation.

\subsection{LLMs for HPC}
LLMs can streamline HPC workflows by automatically generating parallel code, suggesting algorithmic improvements, and assisting with performance tuning. Domain-specific models like OMPGPT~\cite{chen2024OMPGPT} and MonoCoder~\cite{kadosh2024monocoderdomainspecificcodelanguage} demonstrate potential for OpenMP code generation albeit limited to a few clauses while MPIrigen~\cite{schneider2024mpirigenmpicodegeneration} shows that fine-tuning on MPI data can greatly boost MPI code generation. Notably, most prior research has relied on decoder-only, autoregressive models that generate code left-to-right and focus on loop-level translations. In contrast, CodeRosetta~\cite{coderosetta} illustrates that training an encoder-decoder model from scratch can effectively translate code at the function level (e.g., between C++ and CUDA), offering enhanced flexibility and performance.

%% file: sections/3-OMP.tex
\section{\ourtool}
\begin{figure*}[ht]
\centering
  \includegraphics[width=\textwidth]{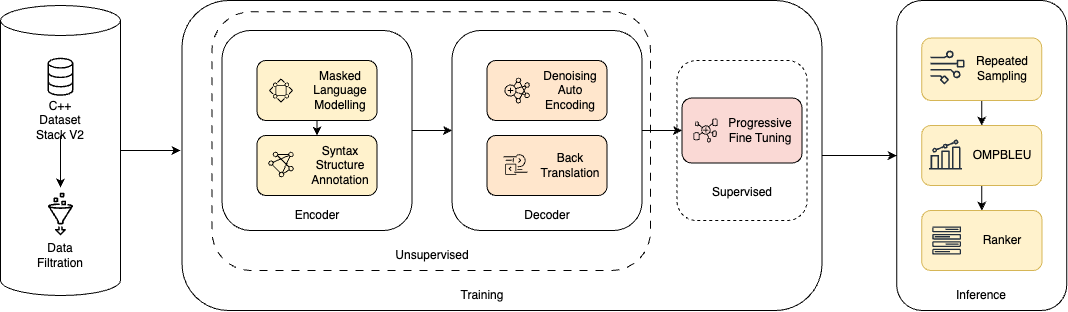}
  \caption{Birds eye view of \ourtool.}
  \label{fig:mlm}
\end{figure*}
\begin{figure*}[ht]
\centering
% Optionally, you can wrap the diagram in a \resizebox if needed:
\resizebox{\textwidth}{!}{
\begin{tikzpicture}[
    font=\sffamily,
    >=latex,
    node distance=2cm,
    every node/.style={align=center},
    block/.style={
        rectangle,
        draw,
        fill=blue!5,
        rounded corners,
        text width=7cm,
        minimum height=1.5cm
    }
]

% --- Input Code Node ---
\node[block] (input) {
    \textbf{Input Code}\\
    \texttt{\#pragma omp parallel for\\
     for (int i = 0; i < n; i += 3) \\
    \quad\quad a[i] = 100;\\
    }
};

% --- Masking Node ---
\node[block, fill=yellow!20, text width=2.8cm,
    right=1.8cm of input] (mask) {
    \textbf{Token\\ Masking}
};

% --- Masked Code Node ---
\node[block, fill=orange!10, right=1.8cm of mask] (masked) {
    \textbf{Masked Code}\\
    \texttt{MASK omp MASK for\\
    for(MASK i = 0; i < n; i += 3) \\
    \quad a[MASK] = 100;\\
    }
};

% --- Cross-Lingual Masked LM Node ---
\node[block, fill=green!10, text width=4cm,
    right=1.8cm of masked] (model) {
    \textbf{\ourtool \\Encoder}
};

% --- Recovered Code Node ---
\node[block, fill=purple!10, right=1.8cm of model] (recovered) {
    \textbf{Recovered Code}\\
    \texttt{\#pragma omp parallel for\\
    for (int i = 0; i < n; i += 3) \\
    \quad a[i] = 100;\\
    }
};

% --- Arrows ---
\draw[->, thick] (input) -- (mask);
\draw[->, thick] (mask) -- (masked);
\draw[->, thick] (masked) -- (model);
\draw[->, thick] (model) -- (recovered);

\end{tikzpicture}

} % end of \resizebox
\caption{Masked Language Modeling}
\label{fig:clm}
\end{figure*}
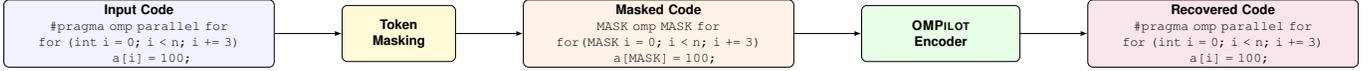
\begin{figure*}[ht]
\centering
\resizebox{\textwidth}{!}{%
\begin{tikzpicture}[
    font=\sffamily,
    >=latex,
    node distance=2.0cm,
    every node/.style={align=center},
    block/.style={
        rectangle,
        draw,
        fill=blue!5,
        rounded corners,
        text width=7cm,
        minimum height=1.4cm
    }
]

% --- 1. Input Code ---
\node[block] (input) {
    \textbf{Input Code}\\
    \texttt{\#pragma omp parallel for reduction(+:sum)\\
    for (int i = 0; i < 10; i++) {
    sum+=i;}}
};

% --- 2. ---
\node[block, fill=yellow!20, text width=2.5cm, right=1.6cm of input] (corrupt) {
    \textbf{Syntax \\Annotation}
};

% --- 3. Annotated Code ---
\node[block, fill=orange!10, right=1.6cm of corrupt] (corrupted) {
    \textbf{Annotated Code}\\
    \texttt{33, 81, 83, 93, 0, 39, 0, 1, 0, 31, 0, 1, 47, 9, 0, 1, 57, 9, 0, 1, 19, 0,}
};

% --- 4. Model (MT Model) ---
\node[block, fill=green!10, text width=3.5cm, right=1.6cm of corrupted] (model) {
    \textbf{\ourtool\\Encoder}
};

% --- 5. Annotated Code ---
\node[block, fill=purple!10, right=1.6cm of model] (recovered) {
    \textbf{Predicted Annotations}\\
    \texttt{33, 81, 83, 93, 0, 39, 0, 1, 0, 31, 0, 1, 47, 9, 0, 1, 57, 9, 0, 1, 19, 0,}
};

% --- Arrows ---
\draw[->, thick] (input) -- (corrupt);
\draw[->, thick] (corrupt) -- (corrupted);
\draw[->, thick] (corrupted) -- (model);
\draw[->, thick] (model) -- (recovered);
\draw[->, thick] (input.east) .. controls +(0,1.5) and +(0,1.8) .. (model.west);

\end{tikzpicture}
} % End of resizebox
\caption{Syntax Structure Annotation}
\label{fig:ssa}
\end{figure*}
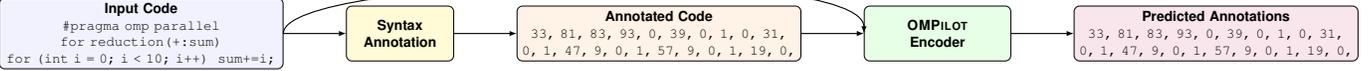
\begin{figure*}[ht]
\centering
\resizebox{\textwidth}{!}{%
\begin{tikzpicture}[
    font=\sffamily,
    >=latex,
    node distance=2.0cm,
    every node/.style={align=center},
    block/.style={
        rectangle,
        draw,
        fill=blue!5,
        rounded corners,
        text width=7cm,
        minimum height=1.4cm
    }
]

% --- 1. Input Code ---
\node[block] (input) {
    \textbf{Input Code}\\
    \texttt{\#pragma omp parallel for reduction(+:sum)\\
    for (int i = 0; i < 10; i++) {
    sum+=i;}}
};

% --- 2. Corrupt Code Node ---
\node[block, fill=yellow!20, text width=2.5cm, right=1.6cm of input] (corrupt) {
    \textbf{Code\\Corruption}
};

% --- 3. Corrupted Code ---
\node[block, fill=orange!10, right=1.6cm of corrupt] (corrupted) {
    \textbf{Corrupted Code}\\
    \texttt{\#pragma omp for MASK(+:sum)\\
    for (MASK i = 0; 10 < i; i+) {
    sum+-i;}}
};

% --- 4. Model (MT Model) ---
\node[block, fill=green!10, text width=3.5cm, right=1.6cm of corrupted] (model) {
    \textbf{\ourtool\\Decoder}
};

% --- 5. Recovered Code ---
\node[block, fill=purple!10, right=1.6cm of model] (recovered) {
    \textbf{Recovered Code}\\
    \texttt{\#pragma omp parallel for reduction(+:sum)\\
    for (int i = 0; i < 10; i++) {
    sum+=i;}}
};

% --- Arrows ---
\draw[->, thick] (input) -- (corrupt);
\draw[->, thick] (corrupt) -- (corrupted);
\draw[->, thick] (corrupted) -- (model);
\draw[->, thick] (model) -- (recovered);

\end{tikzpicture}
} % End of resizebox
\caption{Denoising Auto-Encoding}
\label{fig:denoising-autoenc}
\end{figure*}
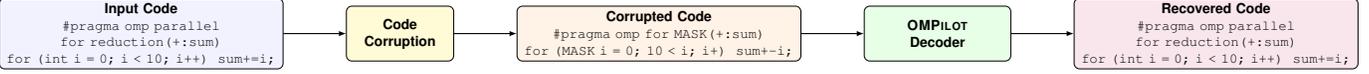
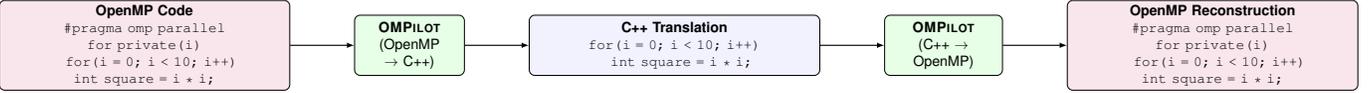
\begin{figure*}[ht]
\centering
\resizebox{\textwidth}{!}{%
\begin{tikzpicture}[
    font=\sffamily,
    >=latex,
    node distance=2.0cm,
    every node/.style={align=center},
    block/.style={
        rectangle,
        draw,
        fill=blue!5,
        rounded corners,
        text width=7cm,
        minimum height=1.5cm
    }
]

% --- 1. Python Code ---
\node[block, fill=purple!10] (pythonCode) {
    \textbf{OpenMP Code}\\
    \texttt{\#pragma omp parallel for private(i)\\
    \quad for(i = 0; i < 10; i++) {\\
        int square = i * i;}}
};

% --- 2. MT Model: Python -> C++ ---
\node[block, fill=green!10, text width=2.5cm,
      right=1.6cm of pythonCode] (model1) {
    \textbf{\ourtool}\\
    (OpenMP $\rightarrow$ C++)
};

% --- 3. C++ Translation ---
\node[block, fill=blue!5, right=1.6cm of model1] (cppCode) {
    \textbf{C++ Translation}\\
    \texttt{for(i = 0; i < 10; i++) {\\
    \quad  int square = i * i;}\\
    }
};

% --- 4. MT Model: C++ -> Python ---
\node[block, fill=green!10, text width=2.7cm,
      right=1.6cm of cppCode] (model2) {
    \textbf{\ourtool}\\
    (C++ $\rightarrow$ OpenMP)
};

% --- 5. Python Reconstruction ---
\node[block, fill=purple!10, right=1.6cm of model2] (pyRecon) {
    \textbf{OpenMP Reconstruction}\\
    \texttt{\#pragma omp parallel for private(i)\\
    \quad for(i = 0; i < 10; i++) {\\
        int square = i * i;}}
};

% --- Arrows ---
\draw[->, thick] (pythonCode) -- (model1);
\draw[->, thick] (model1) -- (cppCode);
\draw[->, thick] (cppCode) -- (model2);
\draw[->, thick] (model2) -- (pyRecon);

\end{tikzpicture}
} % End of \resizebox
\caption{Back-translation}
\label{fig:backtranslation}
\end{figure*}
This section highlights the design and pre-training tasks employed in \ourtool, an encoder-decoder transformer model for OpenMP code translation.
\subsection{Masked Language Modeling}

Masked Language Modeling (MLM) is a self-supervised objective used in transformer architectures such as BERT~\cite{devlin2019bertpretrainingdeepbidirectional} where a percentage of tokens in the input are replaced with a \texttt{[MASK]} token~\cite{lachaux2020unsupervisedtranslationprogramminglanguages}, and the model is trained to predict these masked tokens using their surrounding context. This task helps the model develop a deep understanding of language patterns and semantics, which can be transferred to downstream tasks such as sentiment analysis, question answering, text classification, etc.

In the realm of code, despite unique syntax rules and idioms, the underlying logic remains similar across languages. By learning to predict missing tokens, models improve their ability to generalize across languages, which is particularly valuable when parallel data is scarce. This approach also allows the model to adapt to diverse coding styles, such as varying naming conventions and formatting. Consequently, models such as CodeBERT~\cite{codebert}, GraphCodeBERT~\cite{guo2021graphcodebertpretrainingcoderepresentations}, and CodeT5~\cite{wang2021codet5identifierawareunifiedpretrained,raffel2023exploringlimitstransferlearning} use MLM as a pre-training step before fine-tuning the model on tasks such as code translation, ultimately improving translation accuracy and consistency.

\subsection{Syntax Structure Annotation}
%Unlike general code translation tasks, parallelization requires a deeper understanding of both syntactic and semantic information in the code. To support this, we introduce Syntax Structure Annotation (SSA) as an additional pertraining objective following the MLM stage. SSA leverages Abstract Syntax Trees (ASTs), which offer a hierarchical representation of code that is particularly valuable for reasoning about control flow, data dependencies, and parallelizable regions in code. Such understanding is essential for effective OpenMP parallelization, where correct placement of directives and awareness of loop boundaries and variable scopes are essential.
Parallelization requires a deeper understanding of a program’s syntactic and semantic structure compared to general code translation. To address this, we introduce Syntax Structure Annotation (SSA) as an additional pretraining objective following the MLM stage. SSA leverages Abstract Syntax Trees (ASTs) to provide a hierarchical view of code that is essential for reasoning about control flow, data dependencies, and parallelizable regions, key to correct OpenMP parallelization, which relies on proper directive placement and awareness of loop boundaries and variable scopes. We employ Tree-sitter v0.21.0 \cite{tree_sitter_site_2024} to generate ASTs for each snippet, capturing various syntactic constructs like declarations, expressions, loops, and function definitions. Although OpenMP directives are not explicitly represented in Tree-sitter’s grammar, they are parsed as preprocessor calls: the \texttt{\#pragma} token appears under a \textit{preproc\_call} node, while its arguments (e.g., \texttt{omp parallel for}) are included under a \textit{preproc\_arg} node. We further analyze these subtrees to identify and tag individual OpenMP clauses, ultimately classifying a total of 70 distinct tags.
%We employ Tree-sitter v0.21.0 \cite{max_brunsfeld_2024_10689348} to generate ASTs for each code snippet. Tree-sitter captures various syntactic constructs, such as declarations, expressions, loops, and function definitions, as structured tree nodes. Although OpenMP directives are not explicitly represented in Tree-sitter’s grammar, they are parsed as preprocessor calls: the \texttt{\#pragma} token appears under a \textit{preproc\_call} node, while the remaining components (e.g., \texttt{omp parallel for}) are included in the \textit{preproc\_arg} node. We further analyze this subtree to identify and tag individual OpenMP clauses. We classify a total of 70 different tags.

During pretraining, \ourtool tokenizes the input code and categorizes each token based on its AST node type. Tokens not matched to any AST role are labeled as 0. depicted in Figure~\ref{fig:ssa}. This annotation enriches the model’s understanding of the code structure, facilitating more accurate code generation and improved parallelization outcomes.

% After the MLM stage, we proceed with Syntax Structure Annotation as the next pretraining objective. Abstract Syntax Trees (ASTs) offer a structured representation of code that captures both syntactic and some semantic relationships. We use Tree-sitter v0.21.0 \cite{max_brunsfeld_2024_10689348} to generate ASTs for each code snippet, where each parent node represents a distinct construct—from primitive data types and declarations to expressions and function definitions. Although Tree-sitter does not create specific nodes for OpenMP pragmas, the standard C/C++ grammar interprets \texttt{\#pragma omp} as a preprocessor directive, with \texttt{\#pragma} under the \textit{preproc\_call} node and its arguments under \textit{preproc\_arg}. We extract and tag each clause within the \textit{preproc\_arg} node to cover nearly all OpenMP clauses.

% During pretraining, \ourtool tokenizes the input code and categorizes each token based on its AST role, labeling unmatched tokens as 0 as shown in Figure \ref{fig:ssa}. This approach enhances the model’s understanding of syntactic relationships, which is key to effective code translation and generation.

%\niranjannotes{What is the intuition behind this step? Why do we need it for OpenMP parallelization? What are the categories for tokens? How are they derived? What is multi-modal approach? From the description, it does not look like multi-modal. We should provide an example to demonstrate it.}

\subsection{Denoising Auto Encoding with Weighted Token Cross-Entropy Loss Function}

Denoising Autoencoding (DAE) is a self-supervised learning technique that enhances code translation models by improving robustness and generalization. In DAE, noise, such as token shuffling, character masking, syntax errors, or incomplete code blocks, is deliberately introduced into a code snippet~\cite{lachaux2020unsupervisedtranslationprogramminglanguages}\cite{coderosetta}, and the model is trained to reconstruct the original code. This process helps models better understand programming constructs, variable dependencies, and function relationships, enabling more context-aware translations, even when the input code is incomplete or poorly formatted. This technique is especially valuable given the limited availability of high-quality parallel datasets for code translation for less common languages.

\subsubsection{Training}

To begin DAE training, the decoder is initialized with pre-trained encoder weights. A mix of standard noise injection methods (e.g., random token masking and shuffling) and strategies targeting programming language differences is then applied. This approach increases the chance of removing language-specific keywords, thereby emphasizing critical syntactic components. Additionally, a language-specific token insertion method is used to help \ourtool distinguish between languages and prevent mixing constructs. An adaptive noise strategy is also employed: starting with a low noise ratio and gradually increasing it, so that as training progresses, the model confronts increasingly challenging reconstruction tasks, leading to more resilient representations.

% ---------- Notation ----------

\begin{table}[ht]
\centering
\renewcommand{\arraystretch}{1.15}
\begin{tabular}{@{}ll@{}}
\toprule
Symbol & Meaning \\ \midrule
$B$                 & Batch size (number of code samples) \\
$T_b$               & Length of sample $b$ (after padding) \\
$C$                 & Vocabulary size \\
$p_{b,t,c}$         & Model‑predicted probability that token $(b,t)$ is class $c$ \\
$y_{b,t,c}$         & One‑hot ground‑truth indicator \\
$o_{b,t}\!\in\!\{0,1\}$ & 1 iff token $(b,t)$ is part of an OpenMP construct \\
$m_{b,t}$           & Padding mask (1=real token, 0 = pad) \\
$N=\sum_{b,t} m_{b,t}$ & Effective number of real tokens in the batch \\
$\lambda\!=\!5$   & Weight for OpenMP tokens \\ \bottomrule
\end{tabular}
\caption{Notation used in the loss definitions.}
\end{table}
\begin{equation}
w_{b,t} \;=\;
\begin{cases}
\lambda, & \text{if } o_{b,t}=1,\\[4pt]
1,       & \text{otherwise}.
\end{cases}
\end{equation}

\begin{equation}
\mathcal{L}
= -\frac{1}{N}\,
\sum_{b=1}^{B}\sum_{t=1}^{T_b}
m_{b,t}\,w_{b,t}\,
\sum_{c=1}^{C} y_{b,t,c}\,\log p_{b,t,c}.
\label{eq:wtlf}
\end{equation}

Moreover, we introduce a novel weighted token cross-entropy loss function, as formally presented in Equation \ref{eq:wtlf}, during the pretraining stage. This approach addresses a critical challenge in the training process, namely the significant imbalance between the number of OpenMP-related tokens and the general C++ tokens within the dataset. Given that OpenMP constructs such as directives, clauses, and specific pragma annotations represent a relatively small portion compared to the extensive and varied C++ tokens, standard training procedures inherently underrepresent the learning of OpenMP-specific syntax and semantics.

To counteract this imbalance, we assign higher penalty weights to errors involving OpenMP-related tokens. Specifically, each token is tagged based on its relation to OpenMP constructs (directive keywords such as `parallel`, `for`, `private`, `reduction`, etc.) and assigned a distinct weight during the calculation of the cross-entropy loss. Incorrect predictions involving these tokens thus contribute more significantly to the total loss, incentivizing the model to devote greater learning capacity to accurately predict these critical tokens.

By explicitly focusing the loss function in this manner, we effectively amplify the signal from OpenMP-specific constructs in the overall learning process. This weighted approach ensures the model does not overlook or inadequately learn the nuances and structural correctness required for proper parallelization. Consequently, the training becomes more balanced, significantly improving the model’s proficiency at generating accurate, semantically consistent, and syntactically correct OpenMP parallel code. Our experimental evaluations further validate the effectiveness of this weighted loss strategy, demonstrating notable improvements in the accuracy and consistency of OpenMP pragma placements and clause usage in generated parallel code.

\subsection{Back Translation}

Back translation enhances code translation models by converting a code snippet from one language to another and then back to its original language. This process generates diverse training examples, exposes translation errors, and refines the model. The model is trained in both directions (e.g. C++ to OpenMP and vice versa) by translating batches of source code into a target language and then reconstructing the original source. Comparing the reconstructed code against the original helps identify and correct errors over time. Alternating batches across different language pairs also ensures balanced training and robust translation capabilities.

\subsection{Progressive Fine Tuning}

Progressive fine tuning for OpenMP involves a multi-stage adaptation process in which the pre-trained model is gradually refined using increasingly complex and domain-specific datasets. Initially, the model is fine-tuned on basic code snippets that incorporate standard OpenMP directives, allowing it to grasp essential parallelism constructs. Subsequent phases introduce more intricate examples covering varied clause combinations, nested parallelism, and other advanced patterns, which incrementally enhance the model’s understanding and generalization capabilities. We take the top 15 most occurring clauses for the dataset. We also employ our novel weighted token cross-entropy loss function as shown in Equation\ref{eq:wtlf} during fine tuning. By emphasizing tokens that influence parallelism, our approach enhances both the precision and placement of clauses, ultimately leading to more robust and accurate auto-parallelized code. %\niranjannotes{Why do we need progressive FT? We need to provide intuition behind it. How do you pick examples for this fine-tuning? What is the fine-tuning objective? What is the novel cross-entry loss function - we need to define it. Also, Table 6 should include a row for effect of progressive fine-tuning removal.}

\subsection{Inference}

Sometimes the initial output isn't the best option. Therefore, during inference we generate five different iterations of the same input (score@5), evaluate them using \ourmetric, and then rank the results—selecting the highest-scoring version as our final outcome. %\niranjannotes{This is essentially \texttt{score@5}.}\\addressed

%% file: sections/4-OMPBLEU.tex
\section{\ourmetric}
% This section introduces \ourmetric, a composite metric specifically designed for OpenMP code. This domain-specific focus allows our metric to more accurately assess the parallel semantics and practical correctness of generated code, addressing shortcomings in BLEU and CodeBLEU that can overlook critical aspects of parallel execution.
This section introduces \ourmetric, a composite metric specifically designed for OpenMP code. By focusing on domain-specific characteristics, \ourmetric provides a more accurate assessment of parallel semantics and practical correctness in generated code, addressing limitations of existing metrics such as BLEU and CodeBLEU, which often overlook critical aspects of parallel execution. We first detail each component of \ourmetric in Sections \ref{sec:wc} to \ref{sec:c}, followed by the construction of the composite metric in Section \ref{sec:metric} and its evaluation study in Section \ref{sec:metric_eval}.

\subsection{Weighted Clause Importance Score (WC)}
\label{sec:wc}
This score measures how well the generated code includes key OpenMP clauses (e.g., shared, private, reduction, schedule) compared to the ground truth. It extracts clauses from both versions, assigns weights to each clause, giving higher weights to more critical ones like \texttt{reduction} of 5 and computes the ratio of the weighted overlap between them. A perfect score of 1.0 indicates all expected clauses (weighted by importance) are present, while lower scores reveal omissions or mismatches.\\
Let \(GT\) be the set of clause components from the ground truth and \(Gen\) be from the generated code. With each clause \(c\) in \(GT\) assigned a weight \(w(c)\), the weighted clause score is given by:
\begin{equation}
    WC = \frac{\sum_{c \in GT \cap Gen} w(c)}{\sum_{c \in GT} w(c)}
    \label{eq:wc_equation}
\end{equation}
This metric verifies that all critical parallelism directives and their subcomponents are present, as missing an essential clause (e.g., reduction) can cause incorrect parallel behavior.\\
\lstset{language=C}          % Set your language (you can change the language for each code-block optionally)

\subsection{Variable Usage Consistency Score (VU)}
This score measures the consistency of variable declarations in OpenMP clauses between generated and ground truth code. For each clause (e.g., shared, private, reduction), it extracts the variable sets and computes their Jaccard index. A score closer to 1.0 indicates that the generated code matches the ground truth in identifying variables for parallel behavior.\\
For each clause type \(t\) (e.g., shared, private, reduction, firstprivate, lastprivate), let:
\begin{equation}
    J_t = \frac{\left| V^t_{GT} \cap V^t_{Gen} \right|}{\left| V^t_{GT} \cup V^t_{Gen} \right|}
    \label{eq:intersection_union}
\end{equation}
where \(V^t_{GT}\) and \(V^t_{Gen}\) are the sets of variables extracted from clauses of type \(t\) in the ground truth and generated code, respectively. The overall score is the average over the \(T\) clause types:
\begin{equation}
    VU = \frac{1}{T} \sum_{t \in \{\text{clauses}\}} J_t.
    \label{eq:VU_expression}
\end{equation}
Even if a directive is present, using the wrong or an incomplete set of variables can cause parallel execution errors. This metric quantifies variable consistency by comparing sets, so order doesn't matter (e.g., \texttt{private(i,j)} vs. \texttt{private(j,i)}). However, for clauses like reduction and schedule where order is important, a separate checker is used.

\subsection{Integrated Semantic Similarity Score (IS)}
This metric fuses token-level Levenshtein similarity with embedding-based similarity into a single weighted score. It tokenizes and concatenates all directives to compute a normalized Levenshtein distance yielding 1.0 for textually identical strings. Additionally, it uses a pre-trained model like CodeBERT~\cite{codebert} to derive high-dimensional embeddings for each code version and computes their cosine similarity, capturing semantic likeness beyond mere surface text differences.
Let:
\[
S_{\text{emb}} = \text{cosine similarity between code embeddings,}
\]
\begin{equation}
S_{\text{lev}} = 1 - \frac{D_{\text{lev}}(s_{Gen}, s_{GT})}{\max(|s_{Gen}|, |s_{GT}|)}
\end{equation}
where \(D_{\text{lev}}(s_{Gen},s_{GT})\) is the Levenshtein distance computed on the concatenated OpenMP directive strings \(s_{Gen}\) and \(s_{GT}\)
and \(S_{\text{lev}}\) be the token‑level similarity computed from the normalized Levenshtein distance. Then, with a weighting factor \(\alpha\) (e.g., 0.7),
\begin{equation}
    IS = \alpha\, S_{\text{emb}} + (1-\alpha)\, S_{\text{lev}}.
\end{equation}
This score is robust to minor formatting changes and reordering of tokens while still capturing when the generated code deviates semantically from the ground truth.
\subsection{Ordering Nesting Depth Score (OR)}
This metric verifies that OpenMP directives maintain the correct order, AST nesting level, and collapse clause validity. It leverages tree-sitter to extract directives along with their AST depth. For directives with collapse clauses, it computes actual loop nesting via helper functions and compares it with the declared collapse value, tagging each as "collapse\_valid" or "collapse\_invalid." Finally, it uses difflib to compare the ordered sequence of directives between the generated code and ground truth. Since directive placement is critical for proper parallel execution, any misplacement or incorrect collapse specification will reduce the score.
\subsection{Redundancy and Coverage Score (RC)}
This score measures how well the generated code covers the expected OpenMP directives without adding extras. It uses the Jaccard index to compare the directive sets from the ground truth and generated code, with a penalty applied for any surplus. A high score indicates a close match.\\
Let \(C_{GT}\) and \(C_{Gen}\) be the sets of clause components extracted from the ground truth and generated directives (after normalization). We define the redundancy score as:
\begin{equation}
    R = \frac{|C_{GT} \cap C_{Gen}|}{|C_{GT}|} \times \min\!\left(1, \frac{|C_{GT}|}{|C_{Gen}|}\right).
\end{equation}
This metric penalizes both missing and extra clauses, as unnecessary directives though semantically neutral can add complexity and cause potential performance issues.\\
The semantic similarity score and the redundancy and coverage metric differ in focus and granularity. The semantic similarity score measures overall code similarity using embedding-based comparisons and token-level edit distances, making it robust to minor formatting or ordering variations. In contrast, the redundancy and coverage metric evaluates whether each crucial clause in OpenMP directives is present ignoring irrelevant parts like hardware-dependent num\_threads and penalizes any extra or missing components. Essentially, the former assesses holistic similarity, while the latter offers a detailed check of critical parallel constructs.
\subsection{Cyclomatic Complexity in Parallel Region (CC)}
This score compares the control-flow complexity within OpenMP parallel regions by approximating cyclomatic complexity. It counts decision keywords (e.g., \texttt{if, for, while, case, \&\&, ||}) and adds 1 for each code block, which is extracted using a regular expression. The metric is the ratio of the lower average complexity to the higher average between the generated and ground truth code.\\
Let \(CC_{GT}\) and \(CC_{Gen}\) be the average cyclomatic complexities (where for a code block inside the parallel region,
\begin{equation}
    CC(\text{block}) = (\text{Number of decision points}) + 1
\end{equation}
Then:
\begin{equation}
    CC = \frac{\min(CC_{GT}, CC_{Gen})}{\max(CC_{GT}, CC_{Gen})}
\end{equation}

A large discrepancy in complexity might indicate that the generated code is structurally very different or simplified/overcomplicated, relative to the ground truth.

\subsection{OpenMP Pragma Location Score (PL)} 
This score assesses whether the OpenMP directives are attached to the correct code blocks. In parallel programming, the location of a directive is critical.
\subsubsection{Loop Related Directives}
For "loop-related" directives (those with keywords like "for" or "collapse"), the metric extracts the immediate for‑loop context from the AST and determines the loop's index (its order among for‑loops). It then computes the cosine similarity between the ground truth and generated for‑loop contexts, applying a penalty if the loop indices differ (e.g., a difference of 1 might reduce the score by 50\%).\\
For loop-related pragmas, suppose we extract pairs \((L^i_{GT}, i^i_{GT})\) and \((L^i_{Gen}, i^i_{Gen})\) for the \(i\)th for-loop context (with \(i^i\) being the loop index). Define the cosine similarity for each pair as:
\begin{equation}
    S^i_{\text{loop}} = \cos\big(L^i_{GT}, L^i_{Gen}\big),
\end{equation}
and a loop index penalty:
\begin{equation}
    P^i = \max\!\left(0,\,1 - \frac{\left| i^i_{GT} - i^i_{Gen} \right|}{2}\right).
\end{equation}
Then the overall loop context similarity is:
\begin{equation}
    LS = \frac{1}{n} \sum_{i=1}^{n} \Big( S^i_{\text{loop}} \cdot P^i \Big).
\end{equation}

\subsubsection{Non Loop Related Clauses}
For non‑loop directives such as single, task, taskwait, critical, atomic, barrier or parallel directives not followed by a loop, the metric extracts the immediate context (which may be a compound statement, function call, or expression statement) and computes the cosine similarity between the contexts in the ground truth and generated code.\\
For non-loop pragmas, let \(S_{\text{nonloop}}\) be the average cosine similarity computed over the contexts following non-loop directives. Then, the integrated pragma location metric is:
\begin{equation}
    PL = 
\begin{cases}
S_{\text{nonloop}}, & \text{if no loop contexts are available,} \\[1mm]
\frac{LS + S_{\text{nonloop}}}{2}, & \text{otherwise.}
\end{cases}
\end{equation}
This metric verifies that the generated directive is attached to the correct loop or code block. Misplaced directives (e.g., attached to the wrong loop) reduce the cosine similarity and trigger the loop index penalty.
\subsection{Compilation Score (C)}
\label{sec:c}
This metric ensures that the generated code is syntactically correct and can be compiled. We use Clang-19.7.1 as our compiler with the necessary flags for linking.
\begin{equation}
    C(\text{code}) = 
\begin{cases} 
1, & \text{if the code compiles successfully,} \\[1mm]
0, & \text{otherwise.}
\end{cases}
\end{equation}

\subsection{Composite Metric}
\label{sec:metric}
The composite metric in Equation \ref{eq:ombleu} aggregates individual scores,covering semantic correctness, structural ordering, clause presence, and even compilation, into a single overall score that reflects the quality and correctness of the generated OpenMP code relative to the ground truth. We assign larger weights to components that directly determine OpenMP correctness and placement notably pragma location (PL), compilation success (C), and clause importance (WC) because mistakes there change program semantics or break logic, while lower weights go to components that are informative but less failure-critical (e.g. variable consistency, semantic/surface similarity).By weighting and combining these different dimensions, the metric offers a robust, multi-faceted evaluation that can be fine-tuned empirically based on the importance of each aspect. It is important to note that, when evaluating parallelized code using OMPBLEU, the ground truth code should ideally be authored by domain experts to ensure correctness and adherence to best practices in parallel programming.

\begin{equation}
\begin{aligned}
\text{OMPBLEU} = \alpha \times \text{WC} + \beta \times \text{VU} + \gamma \times \text{IS} + \delta \times \text{OR} \\
               \quad + \epsilon \times \text{RC} + \zeta \times \text{CC} + \eta \times \text{PL} + \theta \times \text{C} \\ \text{where}\\
\alpha(0.3) + \beta(0.05) + \gamma(0.10) + \delta(0.05) + \epsilon(0.05) + \\\zeta(0.05) + \eta(0.2) + \theta(0.2) = 1
\end{aligned}
\label{eq:ombleu}
\end{equation}
\subsection{Metric Evaluation and Ablation Study}
\label{sec:metric_eval}
\begin{figure*}[!t]
  \centering
  \begin{minipage}{0.48\textwidth}
    \centering
    \begin{minted}[frame=lines, framesep=1mm, baselinestretch=1.0, bgcolor=white, fontsize=\scriptsize]{c}
GROUND TRUTH: 
   int main(void) {
    long num_steps = 1000000;
    double step, sum = 0.0;
    int i;
    step = 1.0 / (double)num_steps;
    
    #pragma omp parallel for reduction(+:sum) private(i)
    for (i = 0; i < num_steps; i++) {
        double x = (i + 0.5) * step;
        sum += 4.0 / (1.0 + x * x);
    }
    double pi = step * sum;
    printf("Computed pi = %.16f\n", pi);
    return 0;
}
CASE 1: POOR
{FOR LOOP} 
#pragma omp parallel for private(i) - incorrect placement

CASE 2: BAD
{FOR LOOP}
#pragma omp parallel for reduction(+:sum)

CASE 3: BETTER
#pragma omp parallel for private(i) -correct placement
{FOR LOOP}

CASE 4: BEST
#pragma omp parallel for reduction(+:sum)
{FOR LOOP}
    \end{minted}
    \caption{Code with a single OpenMP directive}
  \end{minipage}
  \hfill
  \begin{minipage}{0.48\textwidth}
    \centering
    \begin{minted}[frame=lines, framesep=1mm, baselinestretch=1.0, bgcolor=white, fontsize=\scriptsize]{c}
GROUND TRUTH: 
   int main(void) {
    int n = 10;
    int total = 0;       
    int extra_sum = 0; 

    #pragma omp parallel for collapse(2) reduction(+:total)
    for (int i = 0; i < n; i++) {
        for (int j = 0; j < n; j++) {
            int value = i * n + j;
            total += value;
            #pragma omp critical
            {
                extra_sum += value;
                printf("Thread %d processed indices (%d, %d) with value %d\n",
                       omp_get_thread_num(), i, j, value); }}}
    printf("Final total (using reduction): %d\n", total);
    printf("Final extra_sum (using critical): %d\n", extra_sum);
    return 0;
}
CASE 1: POOR
#pragma omp parallel for collapse(2)
{FOR LOOP}

CASE 2: BAD
#pragma omp parallel for collapse(2) 
{FOR LOOP
    #pragma omp critical {}}
    
CASE 3: BETTER
#pragma omp parallel for reduction(+:total) 
{FOR LOOP}

CASE 4: BEST
#pragma omp parallel for reduction(+:total)
{FOR LOOP
    #pragma omp critical {}}
    \end{minted}
    \caption{Code with multiple OpenMP directives}
  \end{minipage}
  \caption{Metric Evaluation Scenarios}
  \label{fig:ablation}
\end{figure*}

\begin{table*}[]
\resizebox{\textwidth}{!}{%
\begin{tabular}{|c|c|cccccccc|c|c|c|}
\hline
\multicolumn{1}{|l|}{\textbf{Directive}} & \textbf{Case} & \textbf{\begin{tabular}[c]{@{}c@{}}Weighted\\ Clause(WC)\end{tabular}} & \textbf{\begin{tabular}[c]{@{}c@{}}Variable\\ Usage(VU)\end{tabular}} & \textbf{\begin{tabular}[c]{@{}c@{}}Integrated\\ Semantics(IS)\end{tabular}} & \textbf{\begin{tabular}[c]{@{}c@{}}Ordering\\ (OR)\end{tabular}} & \textbf{\begin{tabular}[c]{@{}c@{}}Redundancy\\ (RC)\end{tabular}} & \textbf{\begin{tabular}[c]{@{}c@{}}Cyclomatic\\ Complexity(CC)\end{tabular}} & \textbf{\begin{tabular}[c]{@{}c@{}}Pragma\\ Location(PL)\end{tabular}} & \textbf{\begin{tabular}[c]{@{}c@{}}Compilation\\ (C)\end{tabular}} & \textbf{OMPBLEU} & \textbf{BLEU} & \textbf{CodeBLEU} \\ \hline \hline
\multirow{4}{*}{\textbf{Single}} & \colorbox{red}{\textbf{1}} & 0.16\reddown & 0.8 & 0.90 & 0 & 0.5 & 0\reddown & 0\reddown & 0\reddown & 20.51\reddown & 89.51 & 95.16 \\ \cline{2-13} 
 & \colorbox{YellowOrange}{\textbf{2}} & 0.83\greenup & 0.8 & 0.93 & 0 & 0.5 & 0\reddown & 0\reddown & 0\reddown & 40.86 & 92.61 & 95.40 \\ \cline{2-13} 
 & \colorbox{GreenYellow}{\textbf{3}} & 0.16\reddown & 0.8 & 0.90 & 0 & 0.5 & 1\greenup & 1\greenup & 1\greenup & 65.52 & 93.48 & 99.23 \\ \cline{2-13} 
 & \colorbox{Green}{\textbf{4}} & 0.83\greenup & 0.8 & 0.93 &1\greenup & 1\greenup & 1\greenup & 1\greenup & 1\greenup & 93.36\greenup & 95.89 & 99.23 \\ \hline \hline
\multirow{4}{*}{\textbf{Multiple}} & \colorbox{red}{\textbf{1}} & 0.16\reddown & 0.83 & 0.84 & 0 & 0.5 & 1 & 0.5\reddown & 1 & 55.08\reddown & 95.46 & 93.51 \\ \cline{2-13} 
 & \colorbox{YellowOrange}{\textbf{2}} & 0.16\reddown & 0.83 & 0.92 & 0.5\greenup & 0.5 & 1 & 1\greenup & 1 & 68.42 & 97.51 & 99.68 \\ \cline{2-13} 
 & \colorbox{GreenYellow}{\textbf{3}} & 0.83\greenup & 0.83 & 0.86 & 0 & 0.5 & 1 & 0.5\reddown & 1 & 75.35 & 95.94 & 93.51 \\ \cline{2-13} 
 & \colorbox{Green}{\textbf{4}} & 0.83\greenup & 0.83 & 0.95\greenup & 0.5\greenup & 0.5 & 1 & 1\greenup & 1 & 88.69\greenup & 97.98 & 99.68 \\ \hline
\end{tabular}%
}
\caption{Metric Evaluation and Ablation Study. The color \colorbox{red}{Red} to \colorbox{Green}{Green} shows a transition from poor to best parallelizations. As the quality of cases improves, \ourmetric scores steadily increase, reflecting a more precise capture of clause detection and placement. In contrast, BLEU and CodeBLEU scores remain high.}
\label{tbl:metric}
\end{table*}
We perform a metric evaluation study (Figure~\ref{fig:ablation}) demonstrating that \ourmetric spans a broader range than BLEU and CodeBLEU, which struggle to capture the nuanced semantics of OpenMP code.

\subsubsection*{\textbf{Scenario 1: Missing or Misplaced Clauses}}
The ground truth loop includes both a \texttt{reduction(sum)} and a \texttt{private(i)} clause. Omitting or misplacing \texttt{reduction} corrupts the result.
\begin{itemize}
  \item \textbf{Case 1:} Only \texttt{private} is emitted—and in the wrong location. The code doesnt compile, the missing \texttt{reduction} incurs a heavy penalty (weighted clause score \reddown), and ordering, cyclomatic, pragma‐location, and compilation scores all drop. BLEU and CodeBLEU remain deceptively high.
  \item \textbf{Case 2:} \texttt{reduction(sum)} appears but on the wrong line. The weighted clause score improves (0.83), but the code does not compile.
  \item \textbf{Case 3:} Only \texttt{private}, correctly placed. It compiles, yet \texttt{sum} is updated incorrectly, yielding an invalid result.
  \item \textbf{Case 4:} Only \texttt{reduction},correctly placed. It compiles and gives higher ordering and redundancy as \texttt{i} is a loop variable and implicitly \texttt{private}. 
\end{itemize}

\subsubsection*{\textbf{Scenario 2: Multiple Directives}}
We mix \texttt{collapse},\newline \texttt{reduction(total)}, and an inner \texttt{critical} for \texttt{extra\_sum}.
\begin{itemize}
  \item \textbf{Case 1:} \texttt{collapse} alone: both \texttt{total} and \texttt{extra\_sum} computed incorrectly.
  \item \textbf{Case 2:} \texttt{collapse} + \texttt{critical}: \texttt{extra\_sum} safe, but \texttt{total} still wrong (no reduction).
  \item \textbf{Case 3:} \texttt{reduction(total)} alone: \texttt{total} correct, \texttt{extra\_sum} data race.
  \item \textbf{Case 4:} \texttt{reduction(total)} + \texttt{critical} (no collapse): both values correct.
\end{itemize}

In both scenarios, BLEU and CodeBLEU yield high, non‐definitive scores despite critical semantic failures, whereas \ourmetric clearly penalizes each mistake.

%% file: sections/5-experimental_setup.tex
\section{Experimental Setup}
This section outlines the experimental configuration for our experiments, offering details on the dataset, the chosen models, and the metrics employed for comparison.
\subsection{Training}
The training was conducted on a single node with 4 Nvidia A100 SXM4 80GB GPUs using HuggingFace Transformers v4.48~\cite{wolf-etal-2020-transformers}. \ourtool features 12 layers with 12 attention heads per layer, a hidden dimension of 1536, and a total of 0.8B parameters. It uses a pre-trained BPE tokenizer from UniXcoder~\cite{guo2022unixcoderunifiedcrossmodalpretraining} (parented on RoBERTa~\cite{liu2019robertarobustlyoptimizedbert}), further trained on our datasets, and employs the GeLU~\cite{hendrycks2023gaussianerrorlinearunits} activation function. The optimizer is AdamW~\cite{loshchilov2019decoupledweightdecayregularization} with a batch size of 16 and gradient accumulation over 2 steps, and the model supports a context length of 512. Masked Language Modeling took 12 hours, Syntax Structure Annotation 2 hours, and Denoising Auto Encoding plus Back Translation 100 hours to train.
\subsection{Dataset}
\subsubsection{Challenge}
A major challenge we encounter is dealing with the limited volume of open source data available. For translation tasks, we require high-quality data in both the source and target languages, which is extremely difficult to obtain.

\subsubsection{Data Preprocessing}
The StackV2~\cite{lozhkov2024starcoder2stackv2} C++ dataset has nearly 76 million files, making random sampling likely to yield low-quality code. To ensure high educational value, we adopted an approach similar to the phi‑1 model~\cite{gunasekar2023textbooksneed}: we randomly selected 100,000 files and used GPT‑3.5 to assign binary labels 'Yes' or 'No'. We then fine‑tune a binary classifier with this data on the CodeSage model~\cite{zhang2024coderepresentationlearningscale} to filter the remaining files. %\niranjannotes{Not sure I understand. Which binary labels?}(addressed)

\subsubsection{Training Set}

\begin{figure}[!ht]
\centering
  \includegraphics[width=60mm]{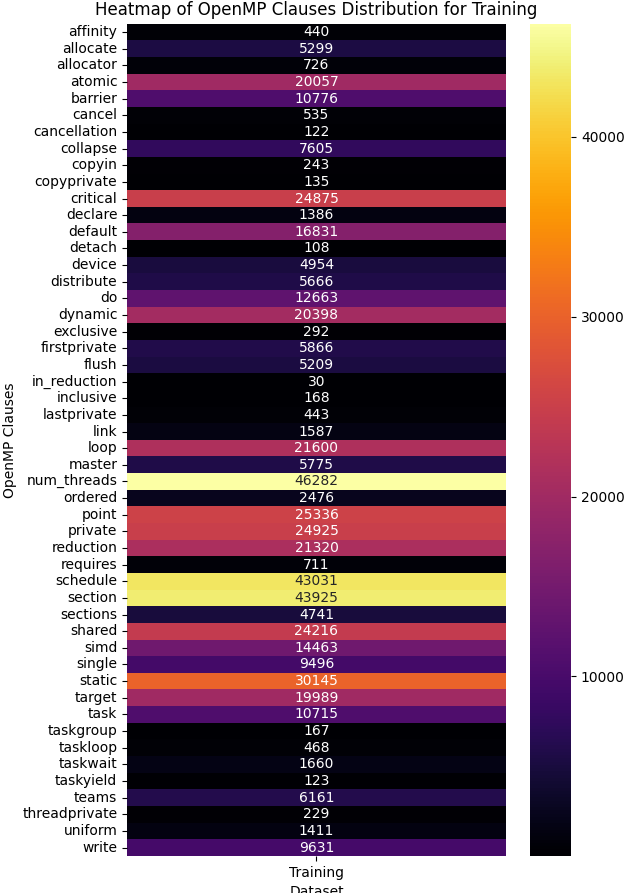}
  \caption{StackV2 OpenMP Training Set Distribution [parallel (56k occurances) \& parallel for (152k occurances) have been ommited for scaling]}
  \label{fig:train}
\end{figure}

We now train our model using the high-quality annotated dataset. We extracted 149,696 unpaired functions containing OpenMP code and paired them with an equal number of C++ functions to avoid bias, yielding nearly 300,000 functions in total. Figure~\ref{fig:train} offers a detailed heatmap of the OpenMP training set distribution by clauses.

\subsubsection{Validation and Test Set}
We use the sets provided by Bin et.al~\cite{lei2023creatingdatasethighperformancecomputing} at LLNL. Our validation set comprises of 182 paired samples while our test set has 26 paired samples.
  
\subsection{Evaluation}
To assess \ourtool's performance in code translation, we report multiple metrics: BLEU and CodeBLEU—widely used in code translation—and ROUGE-L~\cite{lin-2004-rouge}, ChrF~\cite{popovic-2015-chrf}, and METEOR~\cite{meteor} scores. We evaluate \ourtool against the following LLMs: OpenAI o1-mini \& o3-mini~\cite{openai}, Qwen2.5-Coder-14B-Instruct\cite{hui2024qwen2} DeepSeekCoder-V2-Lite-Base~\cite{deepseekai2024deepseekcoderv2breakingbarrierclosedsource}, HPC-Coder-V2-16B which is a finetuned version of DeepSeekCoder-V2-Lite-Base with HPC code on an instruction dataset catered to common HPC topics such as parallelism, optimization, accelerator porting,etc.~\cite{chaturvedi2024hpccoderv2studyingcodellms} StarCoder2-15B-instruct-v0.1~\cite{lozhkov2024starcoder2stackv2}, Codestral-22B-v0.1~\cite{jiang2023mistral7b}
%\begin{itemize}
%    \item \textbf{OpenAI o1-mini \& o3-mini} - Proprietary class leading LLM equipped with reasoning capabilities from OpenAI~\cite{openai} 
%    \item \textbf{DeepSeekCoder-V2-Lite-Base} - Open source 16B parameter model from DeepSeek AI.~\cite{deepseekai2024deepseekcoderv2breakingbarrierclosedsource}
%    \item \textbf{HPC-Coder-V2-16B} - 
%    Finetuned version of DeepSeekCoder-V2-Lite-Base with HPC code on an instruction dataset catered to common HPC topics such as parallelism, optimization, accelerator porting, etc.~\cite{chaturvedi2024hpccoderv2studyingcodellms}
%    \item \textbf{StarCoder2-15B-instruct-v0.1} - Open source 15B parameter model from Bigcode who also bring Stack V2.~\cite{lozhkov2024starcoder2stackv2}
%    \item \textbf{Codestral-22B-v0.1} - Open Source 22B parameter model from Mistral AI.~\cite{jiang2023mistral7b}
    
%\end{itemize}
Inference was carried out on a single Nvidia A100 80GB GPU using the prompt shown in Figure \ref{fig:prompt} for all models. Recognizing that LLMs may not produce optimal results in a single attempt, we sampled five different sequences per query across all models(score@5). %We did not evaluate our model with ParEval~\cite{nichols2024large} because it requires natural language support, which our approach currently does not include.

\begin{figure}
    \centering
\begin{minted}[breaklines,frame=lines,
framesep=1mm,
baselinestretch=1.0,
bgcolor=white,
fontsize=\scriptsize]{c}
You are an expert in translating C++ programs to OpenMP programs.
Given the C++ program below, translate it to OpenMP. Ensure that the OpenMP program is compatible with the C++ program and preserves the semantics of the original code.
Just print the OpenMP program and remove any unnecessary comments.

### C++ Program:{Code from Dataset}

### OpenMP Translation:
\end{minted}
\caption{Prompt for LLM Inference}
\label{fig:prompt}
\end{figure}

%% file: sections/6-results.tex
\section{Results}

In this section, we review the baselines produced by alternative models and compare them to our own. Note that the exact parameter counts in Table\ref{tbl:model} for o1-mini and o3-mini remain undisclosed due to their closed-source status. Additionally, since these models process inputs and prompts via Azure cloud servers, details regarding their on-disk size and power consumption are unavailable, which prevents us from performing local performance profiling. We denote a $\dagger$ to show their times could be network dependent.
\begin{table*}[]
\resizebox{\textwidth}{!}{%
\begin{tabular}{|cc|cccccc|ccc|}
\hline
\multicolumn{2}{|c|}{\textbf{Model}} & \multicolumn{6}{c|}{\textbf{Metrics (Score@5)}} & \multicolumn{3}{c|}{\textbf{Inference}} \\ \hline
\multicolumn{1}{|c|}{\textbf{Type}} & \textbf{\begin{tabular}[c]{@{}c@{}}Parameters\\  (B)\end{tabular}} & \multicolumn{1}{l|}{\textbf{BLEU}} & \multicolumn{1}{l|}{\textbf{CodeBLEU}} & \multicolumn{1}{l|}{\textbf{ROUGE-L}} & \multicolumn{1}{l|}{\textbf{ChrF}} & \multicolumn{1}{l|}
{\textbf{METEOR}} & \multicolumn{1}{l|}
{\textbf{\ourmetric}} & \multicolumn{1}{c|}{\textbf{\begin{tabular}[c]{@{}c@{}}Size on Disk\\ (GB)\end{tabular}}} & \multicolumn{1}{c|}{\textbf{\begin{tabular}[c]{@{}c@{}}Time taken\\ (mins)\end{tabular}}} & \textbf{\begin{tabular}[c]{@{}c@{}}Energy Consumed\\  (Wh)\end{tabular}} \\ \hline
\multicolumn{1}{|c|}{\textbf{\ourtool}} & \textbf{0.8} & \multicolumn{1}{c|}{\textbf{94.38}} & \multicolumn{1}{c|}{\textbf{87.93}} & \multicolumn{1}{c|}{\textbf{93.58}} & \multicolumn{1}{c|}{\textbf{92}} & \multicolumn{1}{c|}{\textbf{92.26}}& \textbf{79.17} & \multicolumn{1}{c|}{\textbf{3.24}} & \multicolumn{1}{c|}{\textbf{0.52}} & \textbf{1.96} \\
\multicolumn{1}{|c|}{o1-mini} & - & \multicolumn{1}{c|}{77.42} & \multicolumn{1}{c|}{70.32} & \multicolumn{1}{c|}{80.7} & \multicolumn{1}{c|}{83.47} & \multicolumn{1}{c|}{74.67} &  70.31 & \multicolumn{1}{c|}{-} & \multicolumn{1}{c|}{9$^{\dagger}$} & - \\
\multicolumn{1}{|c|}{o3-mini} & - & \multicolumn{1}{c|}{86.49} & \multicolumn{1}{c|}{68.70} & \multicolumn{1}{c|}{75.06} & \multicolumn{1}{c|}{86.89} & \multicolumn{1}{c|}{69.42} & 72.23 & \multicolumn{1}{c|}{-} & \multicolumn{1}{c|}{45.3$^{\dagger}$} & - \\
\multicolumn{1}{|c|}{Qwen2.5-Coder} & 14 & \multicolumn{1}{c|}{18.23} & \multicolumn{1}{c|}{34.82} & \multicolumn{1}{c|}{60.74} & \multicolumn{1}{c|}{24.74} & \multicolumn{1}{c|}{55.23} & 69.55 & \multicolumn{1}{c|}{29} & \multicolumn{1}{c|}{31} & 102.93 \\
\multicolumn{1}{|c|}{DeepSeek-CoderV2} & 16 & \multicolumn{1}{c|}{0} & \multicolumn{1}{c|}{1.52} & \multicolumn{1}{c|}{21.35} & \multicolumn{1}{c|}{3.4} & \multicolumn{1}{c|}{25.37} & 11.75$^{*}$ & \multicolumn{1}{c|}{31.4} & \multicolumn{1}{c|}{60} & 90 \\
\multicolumn{1}{|c|}{HPC-CoderV2} & 16 & \multicolumn{1}{c|}{6.56} & \multicolumn{1}{c|}{8.47} & \multicolumn{1}{c|}{68.6} & \multicolumn{1}{c|}{14.05} & \multicolumn{1}{c|}{67.42}& 63.54 & \multicolumn{1}{c|}{35} & \multicolumn{1}{c|}{30} & 51 \\
\multicolumn{1}{|c|}{StarCoder2} & 15 & \multicolumn{1}{c|}{2.97} & \multicolumn{1}{c|}{33.12} & \multicolumn{1}{c|}{68.35} & \multicolumn{1}{c|}{25.19} & \multicolumn{1}{c|}{69.42}& 65.58 & \multicolumn{1}{c|}{32} & \multicolumn{1}{c|}{15} & 73.75 \\
\multicolumn{1}{|c|}{Codestral} & 22 & \multicolumn{1}{c|}{4.32} & \multicolumn{1}{c|}{32.07} & \multicolumn{1}{c|}{56.03} & \multicolumn{1}{c|}{24.56} & \multicolumn{1}{c|}{61.28} & 68.39 & \multicolumn{1}{c|}{45} & \multicolumn{1}{c|}{32.5} & 140.83 \\
\multicolumn{1}{|c|}{OMPGPT} & 0.76 & \multicolumn{1}{c|}{93.52} & \multicolumn{1}{c|}{85.44} & \multicolumn{1}{c|}{93.44} & \multicolumn{1}{c|}{57.57} & \multicolumn{1}{c|}{58.29} & 54.73 & \multicolumn{1}{c|}{-} & \multicolumn{1}{c|}{-} & - \\
 \hline
\end{tabular}%
}
\caption{Model Performance over various metrics and Inference characteristics.}
\label{tbl:model}
\end{table*}
\begin{table}[]
\centering
\resizebox{0.8\linewidth}{!}{%
\begin{tabular}{|c|cccc|c|c|c|}
\hline
\textbf{Model/Metrics} & \textbf{TP} & \textbf{FP} & \textbf{FN} & \textbf{TN} & \textbf{Precision (\%)} & \textbf{Recall (\%)} & \textbf{F1 Score(\%)} \\ \hline
\textbf{\ourtool} & 39 & 13 & 29 & 1323 & \textbf{75} & \textbf{57.35} & \textbf{65} \\
o1-mini & 23 & 39 & 45 & 1297 & 37.09 & 33.82 & 35.38 \\
o3-mini & 29 & 42 & 39 & 1294 & 40.84 & 42.64 & 41.72 \\
Qwen2.5-Coder & 22 & 55 & 46 & 1281 & 28.57 & 32.35 & 30.34\\
DeepSeek-Coder-V2 & 5 & 13 & 63 & 1323 & 27.77 & 7.35 & 11.62 \\
HPC-Coder-V2 & 18 & 21 & 50 & 1315 & 46.15 & 26.4 & 33.64 \\
StarCoder2 & 26 & 23 & 42 & 1313 & 53.06 & 38.23 & 44.44 \\
Codestral & 26 & 52 & 42 & 1284 & 33.33 & 38.23 & 35.61 \\
OMPGPT & 9 & 26 & 59 & 1310 & 25.71 & 13.23 &17.47\\
Intel ICC Classic & 3 & 9 & 65 & 1327 & 25 & 4.41 & 7.50 \\
Cetus & 9 & 3 & 59 & 1333 & 75 & 13.23 & 22.5 \\
 \hline
\end{tabular}%
}
\caption{Classification Results across all models and tools}
\label{tbl:prec}
\end{table}
\begin{figure}[!ht]
\centering
  \includegraphics[width=65mm]{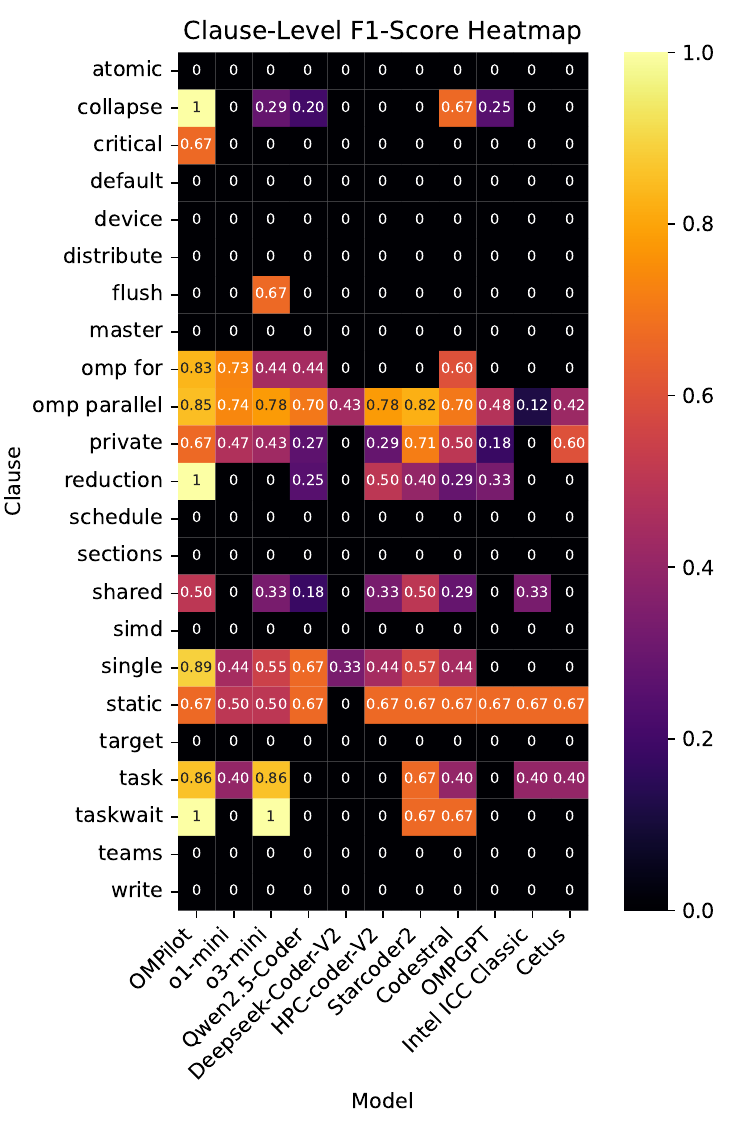}
  \caption{F1 clause level scores ([Higher]Lighter is better). A score of 0 indicates that in those test cases where the clause was present in the ground truth, the model failed to generate the clause. }
  \label{fig:f1}
\end{figure}
\begin{table*}[]
\resizebox{\textwidth}{!}{%
\begin{tabular}{|c|c|c|c|c|c|c|c|}
\hline
\textbf{Test Case} & \textbf{Ground Truth} & \textbf{OMPILOT} & \textbf{o3-mini} & \textbf{StarCoder2} & \textbf{OMPGPT} &\textbf{Intel ICC} & \textbf{Cetus}\\ \hline
\textbf{Case 3} & parallel, single & \textcolor{green}{parallel, single} & \textcolor{green}{parallel, single} & \textcolor{green}{parallel}, \textcolor{red}{section, sections}& - & - & - \\ \hline
\textbf{Case 6} & default, omp parallel for, private, shared & \textcolor{green}{omp parallel for, private} &  \textcolor{green}{omp parallel for,}\textcolor{red}{collapse} & \textcolor{green}{omp parallel for, private}  &  \textcolor{green}{omp parallel for,}\textcolor{red}{collapse}& \begin{tabular}[c]{@{}c@{}}\textcolor{red}{omp parallel, loop}, \textcolor{green}{shared},\textcolor{red}{firstprivate}\end{tabular} & \textcolor{green}{omp parallel for, private} \\ \hline
\textbf{Case 10} & \begin{tabular}[c]{@{}c@{}}atomic, critical, flush,\\ parallel, write\end{tabular} & \textcolor{red}{omp parallel for} & \textcolor{green}{flush, parallel}, \textcolor{red}{shared} & \textcolor{red}{omp parallel for}&- & - & -\\ \hline
\textbf{Case 19} & critical, omp for, parallel & \textcolor{green}{critical, omp for, parallel} & \textcolor{red}{omp parallel for, reduction} & \textcolor{red}{parallel}& \textcolor{red}{omp parallel for, reduction}& - & - \\ \hline
\end{tabular}%
}
\caption{Few Test Cases showcasing Clause Generation}
\label{tbl:case}
\end{table*}
\subsection{Model Performance}
Based on Table \ref{tbl:model}, \ourtool is the leader across all metrics, \textit{BLEU}, \textit{CodeBLEU}, \textit{ROUGE-L}, \textit{ChrF}, and \ourmetric, outperforming models like Codestral (22B parameters) and StarCoderV2 (15B), despite having only 0.8B parameters (3.24 GB on disk). It also excels in efficiency, with the shortest inference time (0.53 minutes) and lowest power consumption (1.95 Wh), while other models require up to 60 minutes and consume significantly more power (e.g., 140.83 Wh for Codestral). Even though o3-mini shows competitive scores (BLEU of 86.49 and CodeBLEU of 68.7), \ourtool surpasses it and all larger models in every metric with a minimal resource footprint. The other models suffer on static metrics due to the presence of unwanted or extra code generations and due to additional whitespaces.
\subsubsection{DeepSeek-Coder-V2 Performance}
Among the evaluated models, DeepSeekCoderV2 had major issues. It failed to generate any code for 16 out of 26 test cases and, when it did produce output, much of it was repetitive newlines, regex commands, or echoes of the prompt, yielding few usable results. This limitation is marked with an asterisk (\(^{*}\)) on \ourmetric in Table \ref{tbl:model}, in contrast to other models that generated code for every test case. Ultimately, DeepSeekCoderV2’s performance illustrates that higher parameter counts alone do not guarantee successful code generation, robust training methods are also crucial.

\subsection{Clause Generation Capabilities}

Our metrics in Table \ref{tbl:prec} evaluate clause-level correctness. For each OpenMP clause from the reference list of all OpenMP clauses, we compute TP, FP, FN, and TN by comparing the model’s output to the ground truth. These are aggregated over all test cases and clauses. This provides a comprehensive measure of how accurately a model detects required clauses (Recall), avoids incorrect ones (Precision), and balances both (F1) with \ourtool excelling in precision, recall, and F1-score.\\
Figure \ref{fig:f1} offers a closer look at clause-level detection, revealing that \ourtool outperforms other models by detecting a higher percentage of clauses consistent with the ground truth. Notably, \ourtool correctly identifies 67\% of all critical clauses, a capability no other model possesses. While \textit{o3-mini} excels at identifying \texttt{flush} clauses, \ourtool is the only model that accurately places the \texttt{reduction} clause. Despite generating multiple \texttt{reduction} clauses (5 by \textit{o1-mini} and 3 by \textit{o3-mini}), these models fail to align them with the appropriate test cases as dictated by the ground truth.
%Table \ref{tbl:case} presents several test cases for detailed analysis among the top-performing open source and closed source models. In Case 3, which contains simple parallel and single directives, both \ourtool and o3-mini correctly generate the directives, whereas StarCoder2 introduces extra clauses (section and sections) and fails to produce the single directive.Case 6 is particularly interesting. Both \ourtool and StarCoder2 generate the correct directives, which, upon closer inspection, yield functionally valid code. In contrast, o3-mini produces an additional collapse clause that is incorrect because it conflicts with an existing dependency between the nested loops, thereby breaking functional correctness. Case 10, involving a more complex function, shows that our model and StarCoder2 are not able to generate the necessary clauses as per the ground truth. o3-mini excels in this case.  Notably, Case 19 highlights \ourtool’s superiority, as it produces a perfect replica of the ground truth, while both o3-mini and StarCoder2 fail to generate the expected clauses.

Table~\ref{tbl:case} provides a detailed analysis of test cases among the top models. In Case 3, featuring simple parallel and single directives, \ourtool and \textit{o3-mini} correctly generate the directives, while \textit{StarCoder2} adds extra clauses and omits the single directive. In Case 6, both \ourtool and \textit{StarCoder2} yield functionally valid code, but o3-mini’s extra collapse clause conflicts with nested loop dependencies, breaking correctness. In Case 10, involving a complex function, \textit{o3-mini} outperforms \ourtool and \textit{StarCoder2} by generating the necessary clauses. Finally, Case 19 underscores \ourtool’s superiority by perfectly replicating the ground truth, unlike the other models.

\subsubsection{Loop Level Parallelism Tool}

OMPGPT is a 0.76B model built on GPT2 that targets loop-level parallelism. The workflow involved extracting loops, generating directives from OMPGPT, and then reintegrating them into the original function. As shown in Table \ref{tbl:prec}, its classification performance is on the lower end, and Figure \ref{fig:f1} further reveals that it misses several directives, resulting in lower scores compared to \ourtool.

\subsubsection{Rule Based Auto Parallelizers}

We evaluate our test set on two rule based auto parallelizers, Intel ICC Classic Compiler 18.2 which supports the \texttt{-parallel} flag and Cetus 2.0~\cite{electronics11050809}. A total of 3 out of the 26 test cases were auto parallelized, with only 1 being correct for ICC. For Cetus, 14 out of the 26 test cases were auto parallelized, with only 4 matching the ground truth. Table \ref{tbl:prec} depicts their classification results.

\subsection{XSBench Reproduction}
% Requires: \usepackage{booktabs}
\begin{table}[t]
\centering

\resizebox{0.9\linewidth}{!}{%
\begin{tabular}{lrrrrr}
\toprule
\textbf{Model} & \textbf{\textsc{\ourmetric}} & \textbf{Clause F1} & \textbf{Speedup@16} & \textbf{Speedup@32} \\
\midrule
\ourtool & \textbf{0.87} & \textbf{0.84} & \textbf{7.1$\times$} & \textbf{12.3$\times$} \\
o3-mini        & 0.80          & 0.78          & 6.3$\times$          & 10.9$\times$          \\
Qwen2.5-Coder  & 0.72          & 0.70          & 5.1$\times$          & 8.7$\times$           \\
\bottomrule
\end{tabular}%
}
\vspace{-0.25em}
\caption{XSBench Each model generates $k{=}5$ candidates; we select top-1 by \ourmetric and report structural/semantic fidelity (\textsc{\ourmetric}, Clause-F1), and runtime speedups versus the serial build under the same toolchain/flags.}
\label{tab:xsbench-reproduce}
\end{table}

\subsubsection{XSBench: Results and Interpretation}
We evaluate XSBench\cite{Tramm:wy} under a reproduce protocol: (i) remove all upstream \texttt{\#pragma omp} directives to form a serial baseline, (ii) have each model propose $k{=}5$ OpenMP variants, (iii) rank by \textsc{\ourmetric} and take the top-1, and (iv) compile and time under identical settings (\texttt{-O3 -fopenmp}). In Table~\ref{tab:xsbench-reproduce}, \textbf{\ourtool} leads on structural/semantic fidelity (\textsc{\ourmetric}~0.87, Clause-F1~0.84) and on throughput (7.1$\times$/12.3$\times$ at 16/32 threads), outperforming \textit{o3-mini} (0.80/0.78; 6.3$\times$/10.9$\times$) and \textit{Qwen2.5-Coder} (0.72/0.70; 5.1$\times$/8.7$\times$). The advantage stems from consistently attaching \texttt{parallel for} to the XSBench hot loop and selecting clauses that mirror expert practice most notably \texttt{reduction(+:tally)} together with \texttt{firstprivate} on loop-carried scalars and a dynamic (or guided) schedule. These choices raise placement and ordering components within \textsc{\ourmetric}, lift Clause-F1 by reducing both clause omissions and redundant privatization, and translate into better scaling on XSBench’s latency/memory-bound kernel.

\subsubsection{Where the Baselines Lose Ground}
\textit{o3-mini} is typically correct on loop targeting but exhibits clause-order variance and conservative extras (e.g., unnecessary \texttt{private(idx)}), which trim precision (Clause-F1) and incur small runtime overheads. \textit{Qwen2.5-Coder} shows larger variance: candidates more often default to \texttt{schedule(static)} on skewed workloads or occasionally omit a needed \texttt{reduction}, depressing both \textsc{OMPBLEU}/Clause-F1 and parallel efficiency. Overall, XSBench illustrates that the expert-aligned structure captured by \textsc{OMPBLEU} proper pragma placement, clause selection, and buildability predicts real performance: the model with the highest structural fidelity is also the one that scales best in practice.

\subsection{Model Ablation}

\begin{table}[t]
\centering

\resizebox{0.9\linewidth}{!}{%
\begin{tabular}{lrrr}
\toprule
\textbf{Experiment} & \textbf{BLEU} & \textbf{CodeBLEU} & \textbf{\ourmetric} \\
\midrule
\textbf{Baseline} & \textbf{94.38} & \textbf{87.93} & \textbf{79.17} \\
\midrule
\begin{tabular}[c]{@{}l@{}}Without Weighted\\ Token Loss Function\end{tabular}
                  & 93.71{\small{} (-0.67\reddown)} 
                  & 87.62{\small{} (-0.31\reddown)} 
                  & 64.89{\small{} (-14.28\reddown)} \\
\begin{tabular}[c]{@{}l@{}}Without SSA\end{tabular}
                  & 93.88{\small{} (-0.5\reddown)} 
                  & 87.71{\small{} (-0.22\reddown)} 
                  & 77.52{\small{} (-1.65\reddown)} \\
\begin{tabular}[c]{@{}l@{}}Without MLM\end{tabular}
                  & 52.35{\small{} (-42.23\reddown)} 
                  & 55.84{\small{} (-32.09\reddown)} 
                  & 11.49{\small{} (-67.68\reddown)} \\
\bottomrule
\end{tabular}%
}
\caption{Model Ablation Study}
\label{tbl:ablation}
\vspace{-0.25em}
\end{table}
Table \ref{tbl:ablation} presents the model ablation results. Removing our weighted token cross-entropy loss slightly reduces BLEU and CodeBLEU scores, but \ourmetric drops by 14 points, indicating its crucial role in detecting additional clauses during inference. This result demonstrates that emphasizing on OpenMP-specific reserved keywords in the pragma directive sharpens our model's focus on crucial parallelization constructs, thereby enhancing its ability to detect and generate the necessary clauses for effective auto parallelization. The sharp decline for MLM reflects its role as the initial pretraining stage, the initialization point for \ourtool.

%% file: sections/7-conclusion_future.tex
\section{Conclusion}

Automatic code parallelization has been extensively explored through static tools (compilers and source-to-source translators) and, more recently, AI-based approaches. Our evaluation, however, revealed two key limitations: AI-based tools rely on natural language, which introduces ambiguity and necessitates large model sizes, and current metrics fail to assess the semantic consistency of OpenMP-based parallel code. To address these, we introduce \ourtool, a domain-specific, efficient transformer-based model for C++ to OpenMP translation, and \ourmetric, a novel metric that leverages syntactic elements of OpenMP pragmas to capture semantic consistency. \ourtool incorporates a unique loss function that emphasizes the syntactic and semantic validity of generated OpenMP code, leading to more accurate parallelization. Experimental results demonstrate that \ourtool outperforms existing LLMs and static tools on multiple metrics, including \ourmetric, and achieves up to 28× higher efficiency by eliminating natural language input. \ourtool also achieves better speedup in real world benchmark codes. Furthermore, our ablation study reveals strengths such as broader support for OpenMP clauses, indicating that \ourtool could drive further research into efficient and accurate auto-parallelization for the high-performance computing community.

%In the future, we plan to deploy the model on architectures using Rotary Positional Embeddings (RoPE)~\cite{su2023roformerenhancedtransformerrotary} to increase context windows without significantly enlarging the model. Additionally, we aim to adopt advanced activation functions like GeGLU or SwiGLU for improved performance, and support systems with Alternating Attention and Flash Attention2~\cite{dao2023flashattention2fasterattentionbetter} to expedite training.